\documentclass[a4paper,11pt]{article}
\pdfoutput=1 

\usepackage{jcappub} 

\usepackage{graphicx}
\usepackage{amsmath}
\usepackage{amssymb}
\usepackage{rotating} 
\usepackage{xspace}
\usepackage{color}
\usepackage{xcolor}
\usepackage{booktabs}
\usepackage[normalem]{ulem}
\usepackage{lscape}
\usepackage{adjustbox}
\usepackage{multirow}

\newcommand{\Msun}{{\rm M}_\odot}

\title{Are there any extragalactic high speed dark matter particles in the Solar neighborhood?}

\author[a]{Isabel Santos-Santos,}
\author[b, c]{Nassim Bozorgnia,}
\author[a]{Azadeh Fattahi,}
\author[d]{and Julio F. Navarro}

\affiliation[a]{Institute for Computational Cosmology, Durham University,\\
South Road, Durham DH1 3LE, UK}
\affiliation[b]{Department of Physics, University of Alberta,\\ 
CCIS 4-181, Edmonton, Alberta T6G 2E1, Canada}
\affiliation[c]{Theoretical Physics Institute, University of Alberta,\\
CCIS 4-181, Edmonton, Alberta T6G 2E1, Canada}
\affiliation[d]{Department of Physics and Astronomy, University of Victoria,\\ Victoria, BC V8P 5C2, Canada}

\abstract{We use the APOSTLE suite of cosmological hydrodynamical simulations of the Local Group to examine the high speed tail of the local dark matter velocity distribution in simulated Milky Way analogues. The velocity distribution in the Solar neighborhood is well approximated by a generalized Maxwellian distribution sharply truncated at a well-defined maximum ``escape" speed. The truncated generalized Maxwellian distribution accurately models the local dark matter velocity distribution of all our Milky Way analogues, with no evidence for any separate extragalactic high-speed components. The local maximum speed is well approximated by the terminal velocity  expected for particles able to reach the Solar neighborhood in a Hubble time from the farthest confines of the Local Group.  This timing constraint means that the local dark matter velocity distribution is unlikely to contain any high-speed particles contributed by the Virgo Supercluster ``envelope", as argued in recent works. Particles in the Solar neighborhood with speeds close to the local maximum speed can reach well outside the virial radius of the Galaxy, and, in that sense, belong to the Local Group envelope posited in earlier work. 
The local manifestation of such envelope is thus not a distinct high-speed component, but rather simply the high-speed tail of the truncated Maxwellian distribution. }

\begin{document}
\maketitle
\flushbottom

\section{Introduction}
\label{sec:intro}

Evidence for the gravitational interaction of dark matter (DM) with ordinary matter exists on various astronomical scales, from galaxies to the largest scale structures in the Universe (see e.g.~refs.~\cite{Bertone:2010zza, Jungman:1995df, Bergstrom00, Bertone05}). If DM consists of new elementary particles, we can search for its non-gravitational interactions with ordinary matter in the laboratory. Direct detection experiments search for DM particles with  mass in the range of $\sim$~[keV -- TeV] in underground detectors, by measuring the small recoil energy of a target nucleus or electron after interacting with a DM particle passing through the detector. Axion searches, on the other hand, directly search for the conversion of light DM ($\sim [10^{-6} - 10^{-3}]$~eV mass) into photons in the detector. A proper understanding of the DM density and kinematics in the Solar neighborhood is needed in order to interpret the results of these searches.

The Standard Halo Model (SHM)~\cite{Drukier:1986tm} is the most commonly adopted model for the DM distribution in the Milky Way (MW). It assumes that DM is distributed in an isotropic and isothermal sphere around the center of the Galaxy, and that the ``local"  DM velocity distribution (i.e., in the Solar neighbourhood) may be modelled as a Maxwell-Boltzmann distribution with a peak speed close to the local circular speed. Recent high resolution simulations of galaxy formation, which include both DM and baryons, find that a Maxwellian distribution fits well the local DM velocity distribution of simulated MW-like halos~\cite{Bozorgnia:2016ogo, Kelso:2016qqj, Sloane:2016kyi,  Bozorgnia:2017brl, Bozorgnia:2019mjk, Poole-McKenzie:2020dbo, Rahimi:2022ymu}. However, simulations show large halo-to-halo scatter, which leads to a large range of best fit peak speeds for the local Maxwellian velocity distribution. 

In addition, simulations studying the effect of large satellite galaxies such as the Large Magellanic Cloud (LMC) on the MW show that the LMC  significantly impacts the high velocity tail of the local DM velocity distribution, shifting it to higher speeds~\cite{SmithOrlik2023, Besla:2019xbx, Donaldson:2021byu}. This  has strong consequences for the interpretation of results from direct detection experiments, leading to considerable shifts of the direct detection exclusion limits towards smaller DM-nucleon or DM-electron cross sections and DM masses. In general, the impact of the LMC on direct detection limits is more significant for low mass DM, where   the experiments probe the high speed tail of the local DM velocity distribution.

Prior work has suggested a further effect that may impact the highest velocities expected for DM particles in the Solar neighborhood, namely the possibility that ``extragalactic" DM particles from the ``envelopes" of the Local Group (LG) or the Virgo Supercluster may add a population
of particles in a narrow range of velocities (about the MW local escape speed for the LG envelope, roughly $600$ km/s, or exceeding $1000$ km/s for the Virgo Supercluster)~\cite{Freese:2001hk,Baushev:2012dm,Herrera:2021puj,Herrera:2023fpq}. Such high speed DM particles in the Solar neighborhood can significantly impact the sensitivity of DM direct detection experiments~\cite{Herrera:2021puj, Herrera:2023fpq}.

If such component exists, it should be present in cosmological simulations of the formation of the LG, which capture the highly dynamic nature of the assembly of the halos of the MW and M31 galaxies and their time-evolving configuration in the proper cosmological context. In this work, we use the APOSTLE hydrodynamical simulations~\cite{Sawala2015, Fattahi2016} of LG volumes to investigate whether a high speed extragalactic DM particle component is found in the Solar neighborhood of simulated MW-like halos. We consider MW analogues in both  dark-matter-only (DMO) and hydrodynamical (Hydro) runs of the APOSTLE simulations at three resolution levels, and study in detail their local DM velocity distributions, with particular emphasis on their high-speed tail. In particular, we focus on what sets the maximum speed of DM particles in the Solar neighborhood, and on whether it may be affected by the presence of extraneous matter of extragalactic origin. 

This paper is structured as follows. In section~\ref{sec:simulations} we provide details on the APOSTLE simulations.  In section~\ref{SecAnal} we present the local DM velocity distribution extracted from the simulated halos, model its high speed tail, and present a physical interpretation for the origin of the maximum speed expected in each halo. We discuss and summarize our main results in section~\ref{sec:conclusions}. 

\section{Simulations}
\label{sec:simulations}

The APOSTLE project~\cite{Sawala2015,Fattahi2016} consists of a suite of cosmological N-body/hydrodynamical simulations of LG-like volumes selected from a $100^3\,$Mpc$^3$ periodic box and resimulated at different resolutions using the zoom-in technique \cite{Jenkins2013}. The LG-like volumes were identified based on the kinematic properties of the MW-Andromeda pair and of the surrounding Hubble flow. More precisely, each volume includes a pair of $\sim 10^{12}\, \Msun$ halos with separation of $[600-1000]$~kpc, and relative radial and tangential velocities of $[-250,0]$~km/s and $[0-100]$~km/s, respectively. The selection also attempts to match, as much as possible, the observed Hubble flow out to $3$ Mpc. Like the observed LG, the halo pairs are relatively isolated, with no other halos of comparable or greater mass within 2.5 Mpc from their barycentre. Details of the LG volume selection process can be found in ref.~\cite{Fattahi2016}. 

The simulations were performed using a modified version of the Tree-PM SPH code, P-Gadget3 \cite{Springel2001a}, with the EAGLE galaxy formation model \cite{Schaye2015}\footnote{using the ``Reference'' model in the language of ref.~\cite{Schaye2015}.} which has been shown to reproduce a wide range of galaxy properties such as stellar mass function, mass-size relation and observed rotation curves \cite{Schaye2015,Crain2015,Schaller_2015}. 
The simulation volumes have also been run in DMO mode, where all the simulation particles are treated as gravitating collisionless fluid. We use both the Hydro and DMO runs in this study. 

Halos and galaxies in the simulations were identified using the Friends-of-Friends (FoF) algorithm with the linking length of 0.2 times the mean interparticle separation, followed by the SUBFIND algorithm to find self-bound halos and subhalos \cite{Springel2001a}. Cosmological parameters of the simulations are adopted from WMAP-7 \cite{Komatsu2011} and are those of a flat $\Lambda$CDM universe with $h=0.704$, $\Omega_0=0.272$, $\Omega_\Lambda=0.728$, $\Omega_{\rm bar}=0.0455$, $\sigma_8=0.81$.

The APOSTLE suite includes 12 LG-like volumes, but for this study we only use 4 volumes that have been simulated at all three resolution levels, both as DMO and Hydro. This enables us to check the sensitivity of our results to numerical limitations. DM and gas particles have masses of $\sim5\times10^4\, \Msun$ and $\sim10^4\, \Msun$, respectively, for the highest resolution (HR) runs. The particle mass increases by $\sim10\times$ and $\sim 100\times$ in the case of medium resolution (MR) and low resolution (LR), respectively, relative to HR\footnote{These resolution levels are labelled as L1/L2/L3 for LR/MR/HR, respectively, in ref.~\cite{Fattahi2016}.}. The gravitational softening length is $134$, $307$, and $711$ kpc for HR, MR, and LR runs, respectively.

Amongst the 8 primary halos in the four APOSTLE volumes studied here (AP-1, AP-4, AP-6, and AP-11), 4 of them have a satellite as massive as the LMC, although they may not match the current position and velocity of the LMC~\cite{SantosSantos2021,SmithOrlik2023}. There has been no attempt in APOSTLE to match other observational constraints on larger scales, like the presence of nearby groups such as M81 and Cen A; the Virgo cluster; or the Supergalactic Plane. As we discuss below, we do not believe that this omission should have a substantial impact on our conclusions.

\section{Analysis and Results}
\label{SecAnal}

\subsection{Local dark  matter velocity distribution in APOSTLE halos}
\label{sec:velocity dist}

\begin{figure}[t]
\centering
\includegraphics[width=0.5\linewidth]{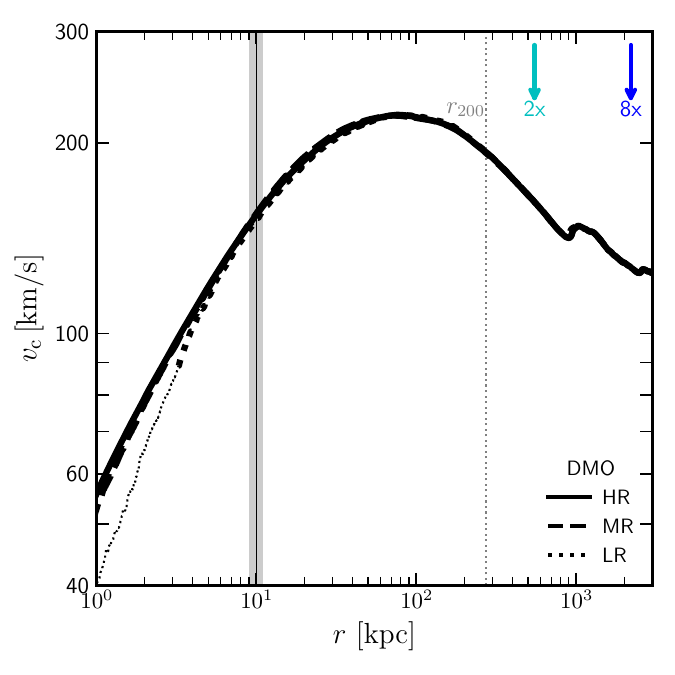}
\caption{Circular velocity profile of a DMO APOSTLE halo shown for the three different resolution levels. HR, MR, and LR are shown in thick solid, dashed, and dotted line types, respectively. 
Note the excellent numerical convergence between the three different runs. 
The LR line becomes thinner at radii where the difference with HR is greater than 10\%.  The gray vertical band indicates the range of ``Solar neighborhood" radii considered when extracting the ``local'' DM velocity distributions. The circular velocity at that radius shows excellent convergence for all three resolutions.  The vertical dotted lines indicate the virial radius, $r_{200}$. Downward pointing arrows mark $2\times$ and $8\times r_{200}$.
}
\label{FigVcProf}
\end{figure}

We select at random one of the eight primary halos in the four APOSTLE volumes to illustrate our analysis of the local DM velocity distribution. This halo is the most massive of the primary pair in APOSTLE AP-6. The $z=0$ circular velocity profiles, $v_c(r)=\sqrt{GM(<r)/r}$, of this halo are shown in figure~\ref{FigVcProf} for the DMO runs at three resolution levels (HR in thick solid, MR in dashed, and LR in dotted line types). Assuming spherical symmetry, circular velocities only depend on the total enclosed mass, $M(<r)$, within radius $r$, measured from the halo centre, which is identified with the position of the gravitational potential minimum. 

Note the excellent agreement between the LR, MR, and HR runs, except  for the innermost regions of the lowest resolution run, which shows a small deficit of mass due to the limited number of particles and the use of a gravitational softening. Outside $0.6$ kpc (for MR) and $3.3$ kpc (for LR) (marked by the transition from thin to thick linetype in figure~\ref{FigVcProf}) all profiles agree very well. The gray vertical shaded area indicates the position of the ``Solar neighborhood" in the simulations, defined as a spherical shell with an inner and outer radius of 9 kpc and 11 kpc, respectively. The vertical dotted line indicates the virial\footnote{Virial quantities are computed at a radius encompassing a mean density contrast of $200$ times the critical density of the Universe, and are identified by a ``200" subscript.} radius, $r_{200}$, of the halo. Downward pointing arrows indicate radii equal to $2\times$ or $8\times$ the virial radius, for future reference. All three simulations show good convergence for the circular velocity in the Solar neighborhood. The number of particles in the shell vary from $2,214$ for the LR halo to $24,946$ for MR, and $37,560,165$ for HR.

\begin{figure}[t]
\centering
\includegraphics[width=1\linewidth]{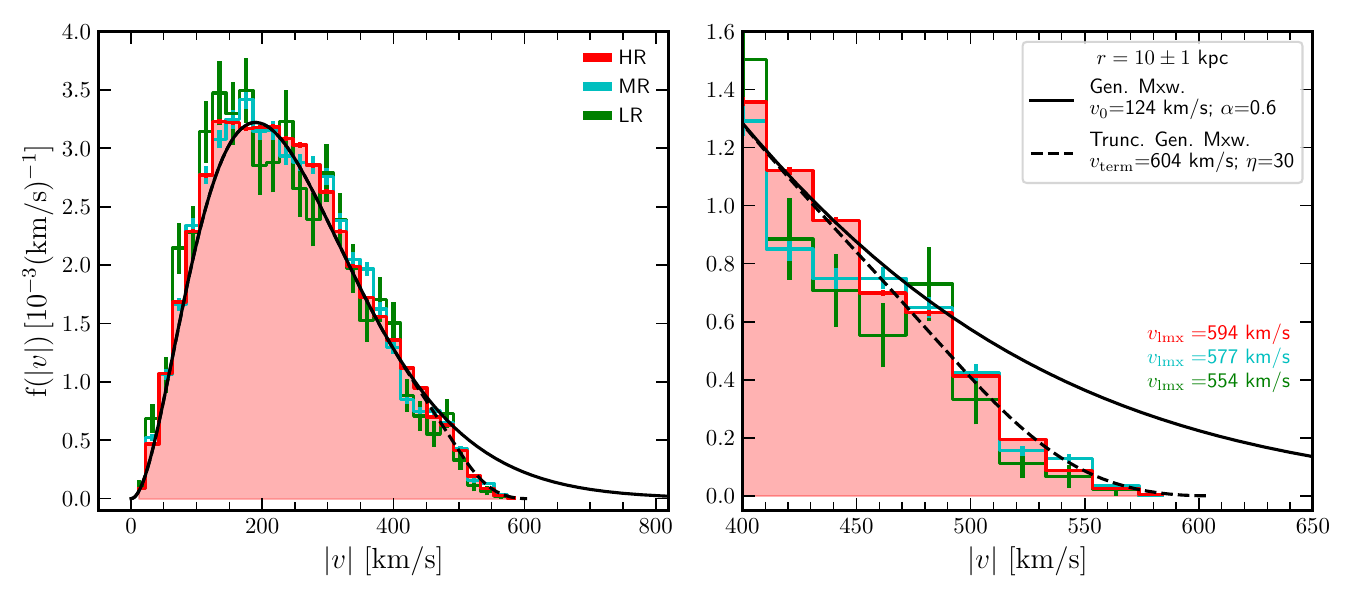}
\caption{Speed distribution of DM particles in the Solar neighborhood of the same DMO halo  shown in figure~\ref{FigVcProf}. The Solar neighborhood is defined as   a spherical shell at $r=10\pm 1$ kpc. The  full range of DM speeds  is shown on the left panel; the right panel zooms in on the high speed tail of the distribution. The red, cyan, and green histograms show the distributions for the HR, MR, and LR levels, respectively, with $1\sigma$ Poisson error bars. The maximum DM particle speed, $v_{\rm lmx}$, reached in each case is quoted in the legend. The solid and dashed black lines show the generalized Maxwellian (eq.~\eqref{EqGMxw}) and the \emph{truncated} generalized Maxwellian (eq.~\eqref{EqTGMxw}) models that best fit the peak and shape of the local DM speed distribution for the HR level, with parameters as listed on the right panel legend.
}
\label{fig:simmodel}
\end{figure}

Figure~\ref{fig:simmodel} shows the speed distribution of DM particles in the Solar neighborhood for the same halo shown in the previous figure. The left panel shows the full range of  DM particle speeds, whereas the right panel zooms in on the high speed tail of the distribution. Velocities are measured in the halo-centric rest frame
(computed by SUBFIND as the center-of-mass velocities of the halo DM particles),
and the velocity modulus distribution (i.e.~speed distribution), $f(|\bf v|)$, is normalized to one, such that $\int f(|{\bf v}|) \, dv=1$.
Histograms of different colors correspond to different resolution levels in the simulations. The error bars represent the $1\sigma$ Poisson uncertainties on the data points.

There is clearly excellent agreement between the three different resolution runs; the best-fit velocity distribution peaks at $190$ km/s and decays rapidly at high speeds, vanishing beyond a well-defined maximum speed of roughly $\sim 600$~km/s. More precisely, the local maximum DM particle speed is $v_{\rm lmx}=554$, $577$, and $594$~km/s for the LR, MR, and HR resolution levels, respectively.

\subsection{The truncated generalized Maxwellian distribution}
\label{SecTGenMxw}

As shown in ref.~\cite{Bozorgnia:2016ogo}, a generalized Maxwellian distribution,
\begin{equation}
 f(|{\bf v}|) \propto |{\bf v}|^2 \exp\left[-(|{\bf v}|/v_0)^{2\alpha}\right],  
\label{EqGMxw}
\end{equation}
 with free parameters $v_0$ and $\alpha$, reproduces the local DM speed distribution of simulated MW analogues quite well. Notice that $\alpha=1$ recovers the standard Maxwellian form. The solid black curve in the left panel of figure~\ref{fig:simmodel} shows that a generalized Maxwellian distribution with $v_0=124$~km/s and $\alpha=0.6$ fits the bulk of the local DM speed distributions from the simulations quite well, except for the high-speed tail, where the simulated distributions show a much sharper truncation than expected from the generalized Maxwellian model.

One important point to note is that there is no hint of a separate ``extragalactic" component contributing an excess of high-speed DM particles; if anything, it is the opposite: there are a lot fewer high-speed particles in the simulated distribution than expected from the tail of the generalized Maxwellian. Although we have shown this in detail for only one APOSTLE halo  in figure~\ref{fig:simmodel}, this is a general result that applies to all the $8$ APOSTLE halos we have analyzed, run either in DMO or Hydro mode.

The differences between the simulated local DM speed distributions and the generalized Maxwellian fit at high-speeds can be better appreciated in the right panel of figure~\ref{fig:simmodel}: there is a clearly defined maximum speed of about $600$ km/s for DM particles in this region, regardless of resolution. Accurate fits to the local DM speed distribution, therefore, require that the generalized Maxwellian model be modified to include a maximum speed. The dashed black line in figure~\ref{fig:simmodel} corresponds to eq.~\eqref{EqGMxw} multiplied by a \emph{correction factor}, 
\begin{equation}
 g(\lvert \textrm{\bf{v}}\rvert) =\textrm{tanh} \left( \eta \left[1-\frac{|v|}{v_{\rm term}}\right]^2  \right).  
\label{EqTGMxw}
\end{equation}
Here $v_{\rm term}$ corresponds to the maximum (``terminal'') velocity beyond which the distribution vanishes, and $\eta$ characterizes the range of speeds over which the correction to eq.~\eqref{EqGMxw} needs to be applied. For the case shown in figure~\ref{fig:simmodel}, $v_{\rm term}=604$~km/s. The simulated distributions start deviating from the generalized Maxwellian at $\sim 400$ km/s (see the right panel of figure~\ref{fig:simmodel}), which implies $\eta=30$. Below $400$ km/s the correction factor (eq.~\eqref{EqTGMxw}) becomes negligible.

\begin{figure}[t]
\centering
\includegraphics[width=0.5\linewidth]{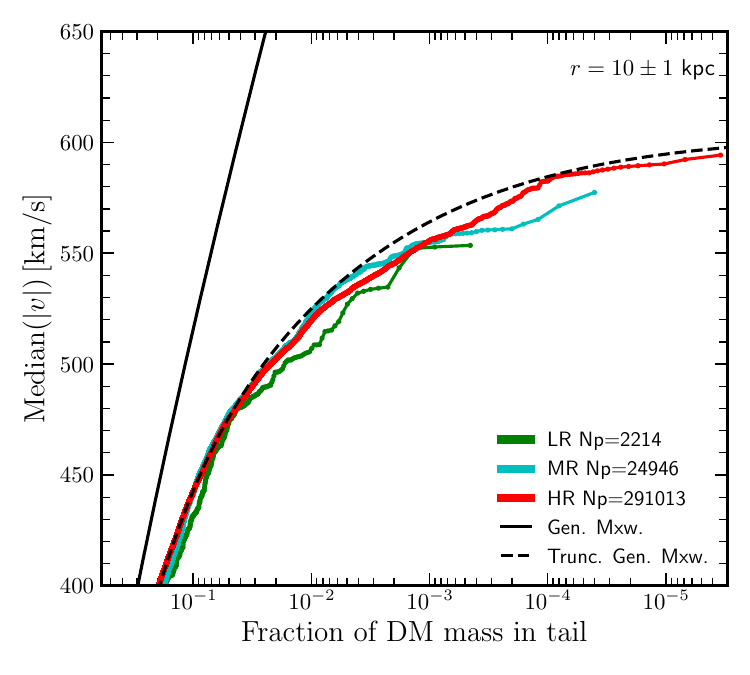}\\
\includegraphics[width=0.9\linewidth]{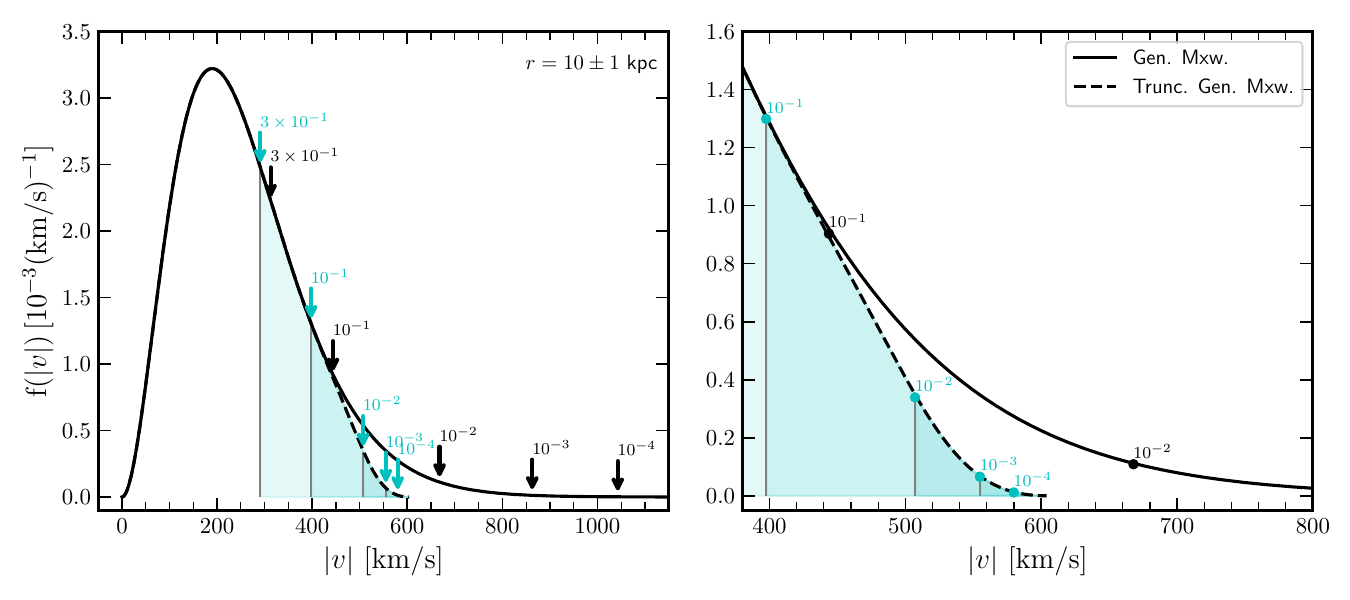}
\caption{Top: the median speed of DM particles in the tail of the local velocity distribution, shown as a function of the fraction of mass in the tail, for the same  halo as in figure~\ref{fig:simmodel}. As in that figure,  HR is shown in red, MR  in cyan, and LR in green. The total number of DM particles in the Solar neighborhood shell ($r=10\pm 1$~kpc) is quoted in the legend for each resolution level. The solid and dashed black lines show the expected median speeds for the generalized Maxwellian and for the truncated generalized Maxwellian fits shown in figure~\ref{fig:simmodel}, respectively.  Bottom left: Best fit generalized Maxwellian (solid) and truncated generalized Maxwellian (dashed) distributions shown in figure~\ref{fig:simmodel}. Downward pointing arrows indicate the location of the tails of the distribution for various mass fractions in the tail, as labelled. Bottom right: a zoomed-in version of the bottom left panel highlighting the high-speed tail.}
\label{fig:fraction}
\end{figure}

\subsection{Local maximum speed and numerical resolution}
\label{SecNumRes}

Is this maximum speed just the result of limited numerical resolution, or is it a physically meaningful local characteristic of the halo? In support of the latter interpretation, we see from the legends in figure~\ref{fig:simmodel} that maximum speed estimates ($v_{\rm lmx}$) increase only slightly with resolution,  by roughly $10\%$ as the number of particles is increased by a factor of $100$. This is about what is expected for a velocity distribution which is actually  truncated at around $\sim 600$ km/s and sampled with different particle numbers.

We show this in the top panel of figure~\ref{fig:fraction}, where we plot the median speed of particles in the high-speed tail of the local DM velocity distribution as a function of the  fraction of the mass in the tail, decreasing from left to right. For the generalized Maxwellian fit to the data in figure~\ref{fig:simmodel}, this median speed is expected to increase without limit as the mass fraction in the tail decreases. From the solid black curves in the top and bottom panels of figure~\ref{fig:fraction} we see that the median speed is expected to increase from $444$ to $668$ to $863$ km/s as the high-speed mass fraction goes from $10\%$ to $1\%$ to $0.1\%$ of the total.

This is clearly {\it not} what is seen in the simulations; the median speed of high-speed particles grow only from $397$ to $507$ to $555$ km/s for the same mass fractions as above, following the increase expected for the truncated Maxwellian distribution shown by the dashed lines in figures~\ref{fig:simmodel} and ~\ref{fig:fraction}. Notice that we use median speeds here rather than the maximum speed to reduce numerical noise in the estimate. The good agreement between the three colored curves obtained from the simulation data at  the three different resolutions and the truncated Maxwellian distribution, thus, strongly suggests that a maximum speed of $\sim 600$ km/s is a true physical property of this APOSTLE halo in the Solar neighborhood.

\subsection{The origin of the local maximum speed}
\label{SecMxVOrig}

The next question that we would like to address is what sets this maximum speed. For an equilibrium isolated system of finite mass, the maximum speed could be safely identified with the local ``escape speed"; i.e., the minimum speed needed for a particle to reach infinity: higher-speed particles would simply leave the system to never return. Unfortunately, DM halos in a cosmological context are neither isolated nor fully in equilibrium, but rather constantly evolving through accretion and mergers. In addition,  total masses in cosmology are not well defined, as enclosed masses increase without bound with increasing radius, confounding the concept of escape speed.

Maximum speeds in non-linear objects formed by accretion (also referred to as ``secondary infall" in the literature) result from the depth of the potential well as well as from the finite age of the Universe, which restricts the maximum radius from where a particle could have been accreted by the present time \cite{bertschinger1983}. Particles coming from even further away could in principle reach higher speeds, but they have not yet had time to reach a particular location. The finite age of the Universe then imprints a maximum turnaround radius (and thus a maximum speed) for particles at a given radius.

We can use this idea to provide a simple estimate of the maximum speeds expected at any radius of a halo by using its present-day mass distribution and the age of the Universe. For simplicity, the estimate assumes that the mass distribution does not evolve in time and is spherically symmetric; this is clearly a rather crude approximation which, however,  works quite well at reproducing the simulation results.

\begin{figure}[t]
\centering
\includegraphics[width=0.5\linewidth]{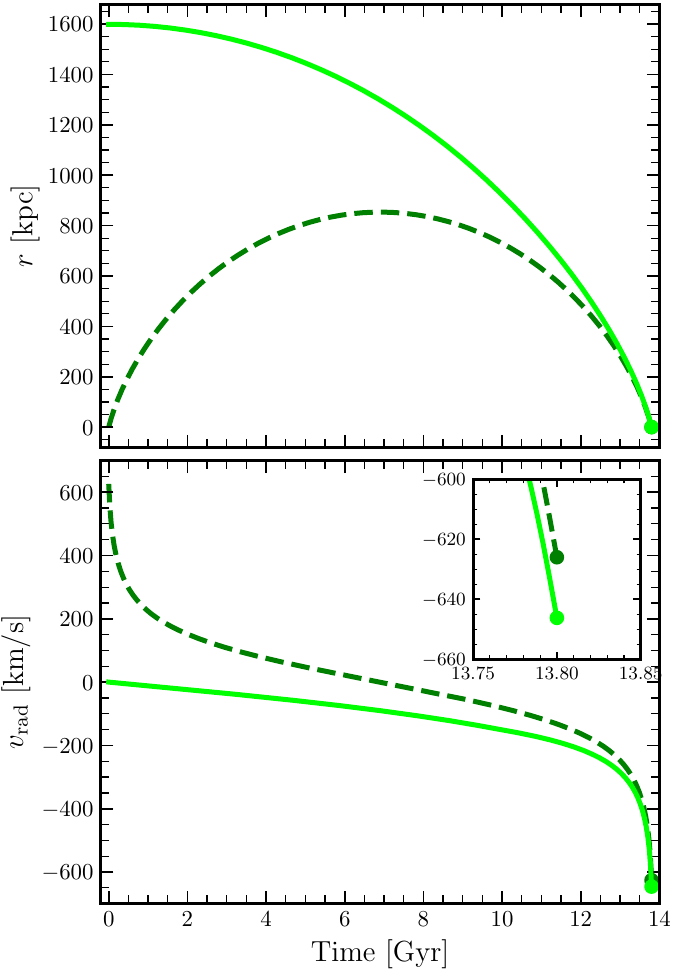}\\
\caption{Radial distance (top panel) and radial velocity (bottom panel) of a test particle as a function of time, orbiting within the gravitational potential of the same DMO halo as in figure~\ref{FigVcProf}. Two radial orbits are shown, both arriving at the halo centre for the first time after $13.8$ Gyr (the age of the Universe). The orbits assume an unevolving spherical Newtonian potential with the same $M(<r)$ profile as the halo and its surroundings at $z=0$. The dashed curve corresponds to an orbit that is initially at $r=0$, the solid line corresponds to an orbit which starts at rest at $t=0$. The circles show the results at redshift $z=0$ in each case.  The inset panel zooms in on the ``terminal velocity" of both orbits at $z=0$ (solid circles).\\}
\label{fig:orb}
\end{figure}

To illustrate the model, figure~\ref{fig:orb} shows the orbit of a test particle on a radial orbit that reaches  the centre of the halo at the present day after evolving for $13.8$~Gyr. The acceleration on the particle is computed from the sphericalized  mass distribution of the halo and its surroundings at $z=0$; i.e., for the same $M(<r)$ profile as in the simulation. Two cases are shown; one (dashed curves) where the particle was initially at $r=0$ and moving radially outwards with enough speed to reach a turnaround radius of $\sim 850$ kpc before returning to the centre after $13.8$ Gyr. The second case (solid green curves) contemplates the case of a particle that starts at rest at $t=0$ from the maximum radius ($r\approx 1.6$ Mpc) from which such particle may reach the centre by the present time. As may be seen from the inset in the bottom panel of figure~\ref{fig:orb}, the terminal velocity  of the particle in case 2 is only slightly higher than in case 1, indicating that much of the acceleration occurs well inside $800$ kpc, the smaller of the two turnaround radii. 

We shall hereafter adopt case 2 to define ``terminal velocities" ($v_{\rm term}$) at any radius of the halo, although we emphasize that adopting case 1 would give similar results. The same model, applied to the Solar neighborhood (i.e., $r=10$ kpc, rather than the halo centre, as in the example shown in figure~\ref{fig:orb}), yields terminal velocities of order $600$ km/s, in excellent agreement with the local maximum speeds obtained in the simulations and with the asymptotic limit shown by the dashed curve in the bottom panel of figure~\ref{fig:fraction}.

\begin{figure}[t]
\centering
\includegraphics[width=0.5\linewidth]{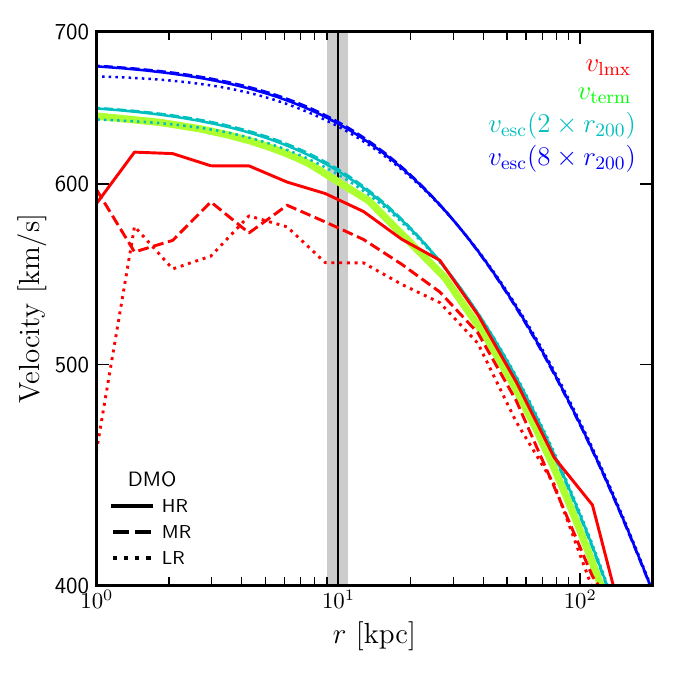}
\caption{Maximum speeds as a function of radius for the same halo selected for analysis in the previous figures. The grey vertical band identifies the location of the Solar neighborhood, as adopted in this study. Red curves indicate the maximum speeds for the HR (solid), MR (dashed), and LR (dotted) runs, computed in concentric spherical shells. Maximum speeds increase slightly with increasing resolution, and approach the terminal velocity curve (shown in solid green) computed using the model illustrated in figure~\ref{fig:orb}. The terminal velocities agree quite well with the Newtonian ``escape speeds" computed assuming that the $z=0$ mass distribution of the halo is truncated at $2\, r_{200}$. Such escape speeds are ill defined in cosmology, and depend on the truncation radius assumed: blue curves show, for example, escape speeds for a truncation radius of $8\, r_{200}$.
}
\label{FigrVmax}
\end{figure}

Actually, the  model reproduces quite well the maximum speeds obtained at {\it all} radii, as may be seen in figure~\ref{FigrVmax}. Here the red curves show the maximum speeds as a function of radius in the halo for the LR (dotted), MR (dashed) and HR (solid) runs. The thick green curve shows the result of the terminal velocity model presented above, which seems to capture well the radial trend of the maximum speed at all radii. Only near the centre does the terminal velocities slightly  overestimate the simulation results, but even there the difference between the predicted maximum speed  and that obtained in the HR simulation is less than $3\%$.

For comparison, we also show in figure~\ref{FigrVmax} the radial dependence of the escape speed from the halo computed by assuming that the gravitational potential of the halo (which we assume is Newtonian) is given by the halo's mass distribution at $z=0$. This distribution needs to be truncated at some radius for the escape speed to converge: we show two cases, truncated at either two (cyan) or eight (blue) virial radii. Interestingly, the terminal velocities are nearly identical to the escape speeds computed using a truncation radius of $2\, r_{200}$. Note that, unlike the escape speed, the terminal velocity model does not require the choice of an arbitrary truncation radius: it only depends on the mass distribution in and around the halo, and on the age of the Universe. In practice, as seen in figure~\ref{fig:orb}, only the matter well inside $\sim 800$ kpc, or inside two to three virial radii seem to really matter.

\subsection{Halo-to-halo scatter}
\label{SecScatter}

Finally, we address how well the terminal velocity model proposed above reproduces the results for other halos in different APOSTLE volumes. Each volume has primary halos of different masses, surrounded by large scale mass distributions that may also vary significantly from one LG realization to another, so this is a rather demanding test of the model.

\begin{figure}[t]
\centering
\includegraphics[width=0.5\linewidth]{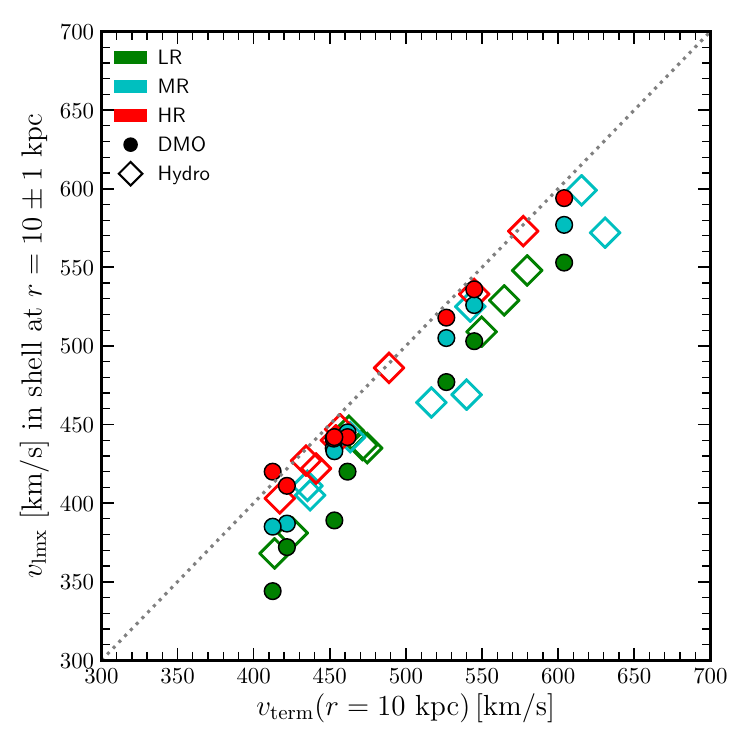}
\caption{Maximum speed of DM particles in the Solar neighborhood (i.e., at  $r=10\pm 1$ kpc)  versus the terminal velocity,  $v_{\rm term}(r=10~{\rm kpc})$, for the eight APOSTLE halos in the DMO (circles) and Hydro (diamonds) simulations at  HR (red), MR (cyan), and LR (green) resolution levels.  Maximum speeds increase slightly but systematically with increasing resolution, and converge to the $1$:$1$ line (in dotted black) at the highest resolution (shown in red). Note the excellent agreement between the terminal velocity model and the local maximum speed for all HR halos at the Solar neighborhood.
}
\label{fig:maxv-all}
\end{figure}

We compare in figure~\ref{fig:maxv-all} the maximum speeds of DM particles in the Solar neighborhood (i.e., at $r=10$ kpc) with the results of the terminal velocity model introduced above. Results are shown for all eight APOSTLE primary halos in 4 different volumes, run at LR, MR, and HR resolutions, both for DMO and Hydro runs. As may be seen in this figure, the terminal velocity model does a remarkable job at accounting for the maximum speeds seen in the HR simulations, be them DMO or Hydro. As discussed in figure~\ref{fig:fraction}, LR and MR runs give systematically (but slightly) lower estimates of the maximum speed because their lower resolutions do not allow them to sample the highest velocities of the distribution.

\begin{figure}[t]
\centering
\includegraphics[width=\linewidth]{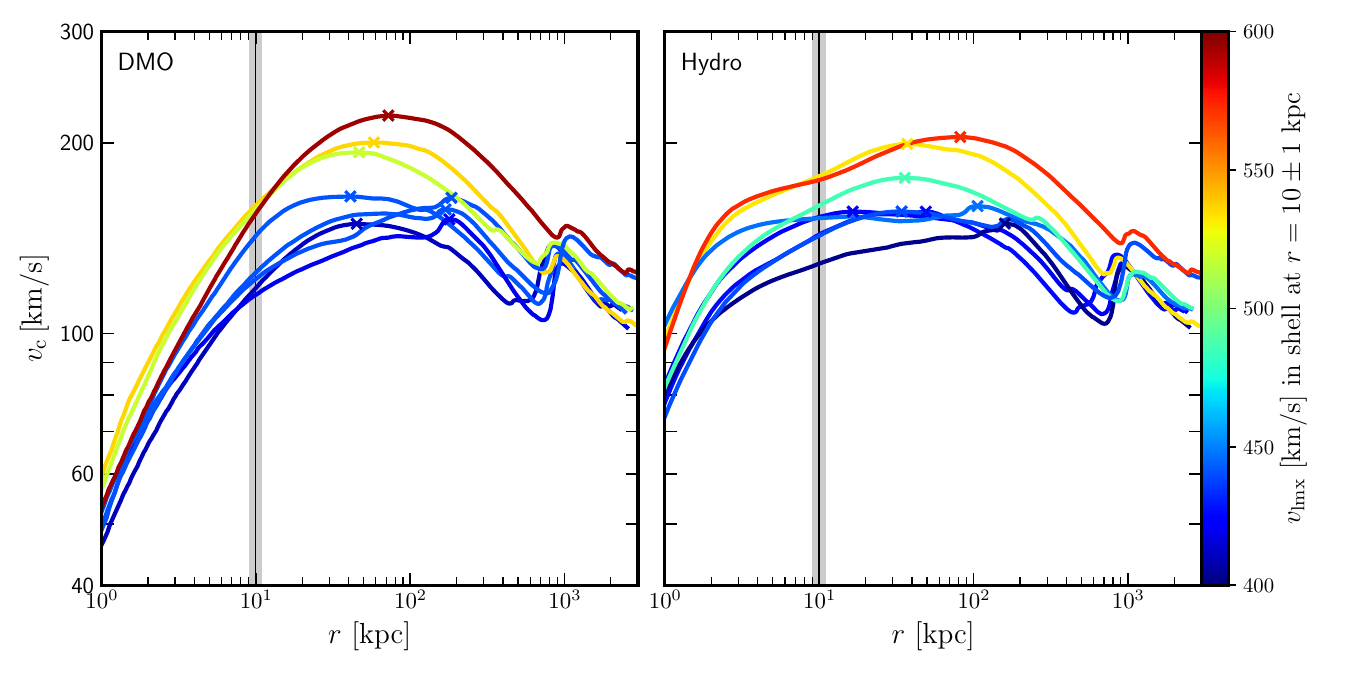}
\caption{Circular velocity profiles, $v_c(r)$, for each of the eight HR APOSTLE halos in the DMO (left) and Hydro (right) simulations. Profiles are colored according to the maximum speed of DM particles in the Solar neighborhood, $v_{\rm lmx}$ (see color bar). Small crosses indicate $V_{\rm max}$, the maximum circular velocity of each system.}
\label{fig:vc-all}
\end{figure}

Figure~\ref{fig:maxv-all} also shows that maximum DM speeds in the Solar neighborhood may vary greatly, even for halos which, like those in APOSTLE, were chosen to be reasonably realistic realizations of MW analogues. The range spans $\sim [350-600]$~km/s, which is much greater than the spread in circular velocities at the same radius, which vary only between $115$ and $159$ km/s for DMO runs, and between $129$ and $177$ km/s in Hydro runs. This may be seen in figure~\ref{fig:vc-all}, where we show the circular velocity profiles of all eight HR APOSTLE halos in the DMO (left panel) and Hydro (right panel) runs. The curves are colored  by their corresponding local DM maximum speed (i.e.~the values on the $y$-axis in figure~\ref{fig:maxv-all}). Maximum circular velocities are marked with small crosses in the figure. 

\begin{figure}[t]
\centering
\includegraphics[width=0.5\linewidth]{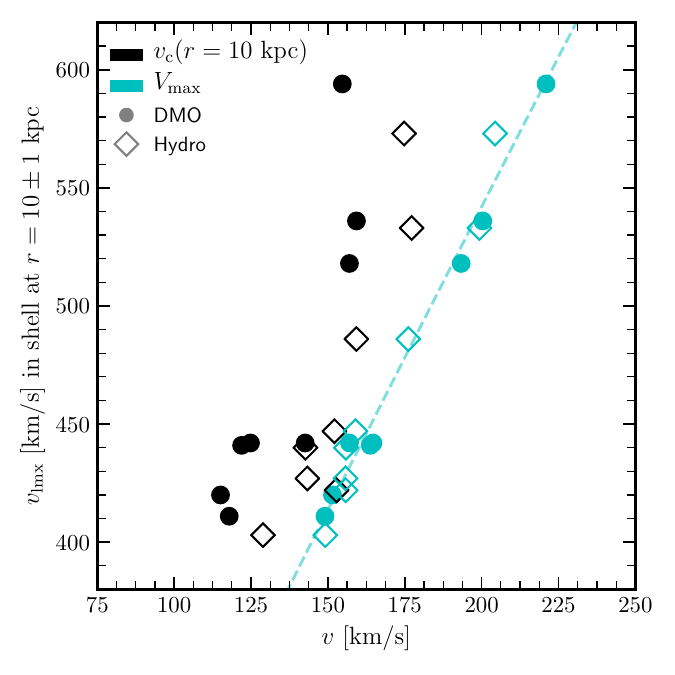}
\caption{Local maximum DM speeds for the DMO (circles) and Hydro (diamonds) halos in the HR run compared with the local circular velocity (i.e., $v_c$ at $10$ kpc; black symbols) and with the maximum circular velocity of each system (cyan symbols). Local maximum DM speeds are poorly correlated with the local circular velocity, but are tightly related to the maximum circular velocity, as shown by the simple proportionality curve, 
$v_{\rm lmx}=3.75 \times V_{\rm max}^{0.94}$, shown in dashed blue.
}
\label{fig:vmaxvmax}
\end{figure}

The correlation between the local maximum DM speed and  circular velocity is thus rather poor, as seen by the black symbols in figure~\ref{fig:vmaxvmax}. This implies that it could be quite difficult to estimate reliably the maximum DM speed expected for the Solar neighborhood using only the circular velocity at the Solar circle as a guide. 

The main reason for this is that terminal velocities depend on the total mass  of the system out to at least twice the virial radius (and beyond), whereas circular velocities depend only on enclosed masses.  Cold DM halos follow Navarro-Frenk-White (NFW) density profiles \citep{NFW1996,NFW1997}, and, for NFW models, the  mass at fixed radii  in the inner regions does not scale linearly with virial mass \cite{Ferrero2021}. For example, a $10^{12}\, \Msun$ NFW halo of average concentration encloses roughly $\sim 4 \times 10^{10}\, \Msun$ within the inner $10$ kpc. Varying the virial mass by a factor of two above and below that value leads to variations of less than $\sim 25\%$ in the DM mass enclosed within that radius, so even small variations in circular velocity may be accompanied by large changes in the terminal velocity at the same radius. 

On the other hand, the correlation between local maximum speeds, $v_{\rm lmx}$, and the maximum circular velocity, $V_{\rm max}$, of the system is excellent, as may be seen from the cyan symbols in figure~\ref{fig:vmaxvmax}. This is mainly because the maximum circular velocities probe better the outskirts of the halos, which are responsible for accelerating infalling particles to the highest speeds. A simple relation, where the maximum speed in the Solar neighborhood is given by $v_{\rm lmx} \sim 3.75\, V_{\rm max}^{0.94}$ fits well the simulation results. The dashed cyan line in figure~\ref{fig:vmaxvmax} shows this relation; for reference, the central escape speed  of an NFW halo is $3.04 \, V_{\rm max}$. Tight constraints on the MW's $V_{\rm max}$ are therefore needed in order to make accurate predictions  about the maximum speed expected for DM particles in the Solar neighborhood.

\section{Summary and Conclusions}
\label{sec:conclusions}

We have used the APOSTLE suite of cosmological hydrodynamical simulations~\cite{Sawala2015,Fattahi2016} to study the high-speed tail of the DM velocity distribution in the Solar neighborhood of MW-like halos. Our main goals are to characterize the tail of the distribution, to determine the origin of the maximum speeds achieved by DM particles, and to search for the presence of distinct components of high-speed DM particles of extragalactic origin hypothesized in earlier work~\cite{Freese:2001hk,Baushev:2012dm,Herrera:2021puj,Herrera:2023fpq}.

In particular, we studied eight APOSTLE  MW  analogues at three resolution levels (LR, MR, and HR), run in dark-matter-only mode, and in Hydro mode; i.e., including gas and galaxy formation. We find that at the location of the Sun (identified with $r=10$ kpc in the simulations) the distribution of DM velocities is well approximated by a generalized Maxwellian distribution (eq.~\eqref{EqGMxw}), sharply truncated at a clearly defined maximum speed (eq.~\eqref{EqTGMxw}). 

There is no sign of a separate ``extragalactic" component, or an excess of high-speed particles over the Maxwellian distribution. If anything, the opposite is true, for all simulations show that the local DM velocity distributions do not extend to arbitrarily large velocities but are truncated at a clearly defined, radially dependent maximum speed. At the Solar neighborhood, this maximum speed varies from halo to halo, but is insensitive to the numerical resolution of the simulations. 

Maximum speeds may be reliably estimated using the $z=0$ mass distribution of the halo (assumed non-evolving) and its surroundings to compute the ``terminal velocity" of a particle in a radial orbit able to reach from rest, in a Hubble time, the Solar circle for the first time. This simple model reproduces remarkably well the results of all APOSTLE DMO and Hydro halos not only at $r=10$ kpc, but at all radii. 

The terminal velocities agree fairly well with the Newtonian ``escape speed" estimated when truncating the $z=0$ halo mass distribution at twice the virial radius. This means that particles near the tail of the distribution have apocentres well beyond the confines of the halo. In that sense, the local high-speed tail does originate in the LG ``envelope" hypothesized in earlier work~\cite{Baushev:2012dm}. However, these envelope particles do {\it not} constitute a discernible separate component of high-speed particles, but are rather just the local representatives of the far outskirts of the halo, which spills beyond its nominal virial boundary. 

We see also no sign of any ``extra-high-speed" DM particles contributed by nearby groups or clusters, a result not unexpected given the success of the terminal velocity model presented above:  particles from outside the LG have not yet had time to reach the Solar neighborhood in the finite age of the Universe. We conclude that it is highly unlikely that the Solar neighborhood contains any $\sim 1000$ km/s extragalactic particles contributed by the Virgo Supercluster, as speculated in previous work~\cite{Baushev:2012dm}. 

The halo-to-halo scatter in the maximum DM speed at the Solar neighborhood is much larger than expected from the scatter in the circular velocity at the same radius. This is because the terminal velocities depend on the mass distribution extending well outside the virial radius, whereas circular velocities depend only on the enclosed mass within the Solar circle. This implies that it is not straightforward to estimate local maximum speeds based solely on the local circular velocity. 

The simulations suggest, on the other hand, a strong correlation between the maximum circular velocity of the system and the local maximum DM speed. This is because the maximum halo circular velocity is a better probe of the mass distribution on  larger scales, which is needed to make an accurate prediction of the local maximum DM speed. If the maximum circular velocity of the MW is $220$ km/s \cite{Cautun:2019eaf}, the simulation results suggest a  local maximum DM speed of $3.75 \times 220^{0.94}=597$ km/s.

This estimate of the local maximum DM speed depends critically on the maximum circular velocity assumed, as well as on the assumption that the Solar neighborhood is close to equilibrium. Strong transients, such as the recent arrival of the Magellanic Clouds into the Milky Way halo, could result in large fluctuations in the halo, causing significant shifts in the high-speed tail of the local DM velocity distribution that have been recently studied in both idealized and cosmological simulations~\cite{SmithOrlik2023, Besla:2019xbx, Donaldson:2021byu}.


\acknowledgments
ISS acknowledges support from the European Research Council (ERC) through Advanced Investigator grant to C.S. Frenk, DMIDAS (GA 786910). NB acknowledges the support of the Canada Research Chairs Program, the  Natural Sciences and Engineering Research Council of Canada (NSERC), funding reference number RGPIN-2020-07138, and the NSERC Discovery Launch Supplement, DGECR-2020-00231. AF is supported by a UKRI Future Leaders Fellowship (grant no MR/T042362/1). This work used the DiRAC Memory Intensive system at Durham University, operated by ICC on behalf of the STFC DiRAC HPC Facility (www.dirac.ac.uk). This equipment was funded by BIS National E-infrastructure capital grant ST/K00042X/1, STFC capital grant ST/H008519/1, and STFC DiRAC Operations grant ST/K003267/1 and Durham University. DiRAC is part of the National E-Infrastructure.

\clearpage

\bibliographystyle{JHEP}
\bibliography{refs}

\end{document}